\def\ee{e^+e^-}
\def\chp#1{\widetilde\chi^{~+}_{#1}}
\def\chm#1{\widetilde\chi^{~-}_{#1}}
\def\chpm#1{\widetilde\chi^{~\pm}_{#1}}
\def\chmp#1{\widetilde\chi^{~\mp}_{#1}}
\def\chz#1{\widetilde\chi^{~0}_{#1}}

\def\selpl{\widetilde e^{~+}_L}
\def\selml{\widetilde e^{~-}_L}
\def\selpr{\widetilde e^{~+}_R}
\def\selmr{\widetilde e^{~-}_R}
\def\sell{\widetilde e_L}
\def\selr{\widetilde e_R}

\def\smupl{\widetilde \mu^{~+}_L}
\def\smuml{\widetilde \mu^{~-}_L}
\def\smupr{\widetilde \mu^{~+}_R}
\def\smumr{\widetilde \mu^{~-}_R}
\def\smul{\widetilde \mu_L}
\def\smur{\widetilde \mu_R}

\def\stau#1{\widetilde \tau_{#1}}

\def\snu{\widetilde \nu}
\def\snue{\widetilde \nu_e}
\def\snum{\widetilde \nu_\mu}
\def\snut{\widetilde \nu_\tau}

\def\iab{ab$^{-1}$}
\def\ifb{fb$^{-1}$}

\def\lc{LC}
\def\lhc{LHC}

\def\me{\mbox{$\rlap{\kern0.2em/}E$}}

\documentclass[twocolumn,prd,amsmath,amssymb]{revtex4}

\usepackage{graphicx}
\begin{document}
\bibliographystyle{revtex}


\title{Run Scenarios for the Linear Collider }



\author{M. Battaglia} 
\affiliation{CERN}
\author{J. Barron, M. Dima, L. Hamilton, A. Johnson, U. Nauenberg, 
M. Route, D. Staszak, M. Stolte, 
T. Turner, C. Veeneman}
\affiliation{University of Colorado}
\author{J. Wells} 
\affiliation{University of California, Davis}
\author{J. Butler, H. E. Montgomery} 
\affiliation{Fermi National Accelerator Laboratory}
\author{R. N. Cahn, I. Hinchliffe} 
\affiliation{Lawrence Berkeley National Laboratory}
\author{G. Bernardi} 
\affiliation{LPNHE. Universities of Paris VI and VII}
\author{J. K. Mizukoshi}
\affiliation{University of Hawaii}
\author{G. W. Wilson}
\affiliation{University of Kansas}
\author{G. A. Blair}
\affiliation{Royal Holloway, University of London}
\author{J. Jaros} 
\affiliation{Stanford Linear Accelerator Laboratory}
\author{P. D. Grannis} 
\email{pgrannis@sunysb.edu}
\affiliation{State University of New York at Stony Brook}


\date{October 16, 2001}

\begin{abstract}
Scenarios are developed for runs at a Linear Collider, in the case 
that there is a rich program of new physics.
\end{abstract}

\maketitle


\section{Introduction}
\label{intro}
The physics program of the linear $\ee$  collider \mbox{LC} is 
potentially very extensive, particularly in the case that a
Higgs boson with mass below 300 GeV is found and relatively
low energy scale supersymmetry (SUSY) exists.  
For such a case, we have examined a possible run plan for the  
\lc ~to explore the new states and their masses, and estimated the precision
on measured parameters that can be attained in a reasonable time
span.   

For this study, we have examined a scenario with
a light 
SM-like Higgs boson of 
mass 120 GeV and two minimal supergravity (mSUGRA) models with
many low mass sparticles.   This scenario is conservative; 
with many particles to study there are many desired
operational conditions for the collider (different energies 
and beam polarizations).
We have not assumed that positron polarization is available, again a 
conservative assumption from the point of view of the running time
required.

We have taken the total time for the runs to be that required to
accumulate 1 \iab ~(1000 \ifb ) at 500 GeV.  Based on estimates \cite{smlum} of the 
luminosity that could be delivered by the \lc ~summarized
in Table~\ref{tab:lumprofile}, we estimate that this represents
a program for the first 6 -- 7 years of \lc ~operation.  Such an
estimate is only qualitative and depends more upon the ultimate
luminosity of the accelerator than upon the details of early low 
luminosity during commissioning.

\begin{table}
\caption{\label{tab:lumprofile}
Profile by year of the luminosity accumulation.  The luminosity 
is given in \ifb ~assuming 500 GeV operation. 
}
\begin{ruledtabular}
\begin{tabular}{cccccccc}
Year & 1 & 2 & 3 & 4 & 5 & 6 & 7 \\
$\int {\cal L}dt$ &10&40&100&150&200&250&250\\
\end{tabular}
\end{ruledtabular}
\end{table}

We have chosen two SUSY benchmarks shown in Table~\ref{tab:msugrapt}:
the TESLA TDR RR1 \cite{teslarr1}
and the Snowmass E3 working group \cite{sm2} point, also
known as benchmark SPS1.  They 
provide a rich spectrum of sparticles at relatively low masses. 
The TESLA RR1 scenario has been used for a variety of previous
studies, but is now ruled out by LEP data.  The SPS1
point gives low sparticle masses, but emphasizes decays via
$\tau$'s and thus provides additional experimental challenges.
For both scenarios, \lhc ~ should have discovered SUSY and explored
some of its aspects prior to the \lc ~operation. However, the precision 
measurements available at the LC, and its access to states 
unobservable at the \lhc , will be needed for the full exploration 
of the new physics.

In devising a proposed run plan, one should keep in mind that 
it is entirely possible, even likely, that there is new physics
to be explored that depends on operation of the \lc ~near its
highest energy.  Thus a run plan that devotes excessive operation
at lower energies may be counterproductive.

\begin{table}
\caption{\label{tab:msugrapt}
Minimal Sugra parameters for the two SUSY benchmark points used in
this study. 
}
\begin{ruledtabular}
\begin{tabular}{c|c|c}
~ & TESLA RR1 ~~~~& SPS1 ~~~~\\ \hline
$m_0$ & 100 GeV ~~~~& 100 GeV ~~~~\\
$m_{1/2}$ & 200 GeV ~~~~& 250 GeV ~~~~\\
$\tan\beta$ & 3 ~~~~& 10  ~~~~\\
$A_0$ & 0 GeV ~~~~& 0 GeV ~~~~\\
sgn($\mu$) & $+$ ~~~~& $+$ ~~~~\\
\end{tabular}
\end{ruledtabular}
\end{table}

\section{Run Plan}
\label{runplan}

For any physics scenario, studies of the  Higgs
boson and the top quark will be high priorities for the linear
collider.  The Higgs studies are possible at any energy above the 
associated $Z H$ production threshold but, due to the 
fall of the cross-section with energy, are best optimized not too
far above threshold. 
The top studies
require operation in the vicinity of the $t\overline t$ threshold
at 350 GeV.

If supersymmetry exists, the desired \lc ~energies and electron beam
polarizations depend sensitively on the specific realization
of the supersymmetric sector.  In many cases, the best 
resolution for the sparticle masses is obtained by dedicated
runs near the threshold for producing a specific particle.  
In this case, however,
it is necessary to establish these particle thresholds with some 
accuracy prior to making a scan.   In some cases these may be 
determined from \lhc ~experiments; in others it is necessary to
establish masses from \lc ~operation at its maximum energy through
the use of kinematic end points in SUSY decay \mbox{chains}.  For this
study, we have imagined that the mass estimates must be obtained
from the LC, so the first order of business for SUSY studies
is operation at the full machine energy.  

The luminosity required
for subsequent threshold runs is dictated by the threshold behavior
of the cross-sections (typically proportional to $\beta$ for 
fermion-antifermion ({\it e.g.} gaugino pair) 
production and to $\beta^3$ for boson ({\it e.g.} sfermion) pair production in 
$\ee$ collisions), and by the dependence of cross-sections
on electron beam polarization.  In some cases, we have concluded
that particular sparticle production cross-sections are too small,
or that decay chains yield too few easily reconstructed particles,
to warrant spending time on a dedicated threshold scan.

The suggested run plans for the two assumed SUSY benchmark points
are shown in Tables~\ref{tab:runrr1} and \ref{tab:runsps1}.
In both plans, we assume that special $e^-e^-$ runs are taken
to obtain high precision measurements of the  $\selr$ mass.  For
the SPS1 point, the $\chpm2$ mass is such that the $\chp1 \chm2$
threshold is above the nominal maximum 500 GeV \lc ~energy; to
reach this crucial state in the SPS1 point, the machine would
need to operate at 
580 GeV, a possibility if one trades luminosity for energy by
higher rf loading of the accelerator structures \cite{smlum}.
For the TESLA RR1 benchmark, we have envisioned two energies
with high integrated luminosity from which mass measurements may
be made from kinematic end point studies.  The 320 GeV
run also serves in this scenario for a scan of
$\snu$ thresholds.  Recent work however suggests that the
precision of end point mass measurements is optimized by running
at the highest available energy.  For the SPS1 point, the 
$\snu$ threshold runs are of limited utility in any case (the 
branching ratios
into low background final states are too small), so for
this point we have included only one run at 500 GeV
for end point measurements.

Note that for both run plans, at least two thirds of the 
accumulated luminosity is acquired at an energy within
80\% of the maximum \lc ~energy, so searches for new
phenomena beyond the Higgs and supersymmetry studies posited
here should be possible.

Tables~\ref{tab:runrr1} and \ref{tab:runsps1} 
show the desired electron polarization states for each energy setting.
We assume an $e^-$ beam polarization of 80\% (and no $e^+$
polarization).
In some cases, we suggest an equal luminosity for
left and right polarized electrons, either to  examine backgrounds
or to access different particle states.  For some of the dedicated
threshold runs, we specify the dominant beam polarization that
maximizes the desired reaction rate; in these cases, we imagine
that perhaps 90\% of the data is taken with this preferred
polarization and the remaining $\sim$10\% of the data with the 
other polarization.   The $e^-e^-$ operation assumes that
both beams are polarized as indicated.   

\begin{table*}
\caption{\label{tab:runrr1}
Run allocations for the TESLA RR1 Minimal Sugra parameters. 
}
\begin{ruledtabular}
\begin{tabular}{cccccl}
Beams & Energy & Pol. & $\int {\cal L}dt$ 
  & [$\int {\cal L}dt]_{\rm equiv}$ & ~~~~~~~~Comments \\  \hline
$\ee $ & 500 & L/R & 245 & 245 & Sit at top energy for heavy sparticle
       end point measurements \\ \hline
$\ee $ & 320 & L/R & 160 & 250 & End point measurements for light 
                                  sparticles \\ 
       &     &     &     &     & Scan $\snu$ pair thresholds \\ \hline
$\ee $ & 255 & L/R & 20 & 40 & Scan $\chp1~\chm1$ threshold \\ \hline 
$\ee $ & 265 & R & 20 & 40 & Scan $\smur$ and $\stau1$ pair 
                                  thresholds \\ \hline 
$\ee $ & 310 & L & 20 & 30 & Scan $\sell~\selr$ threshold \\ \hline
$\ee $ & 350 & L/R & 20 & 30 & Scan $t\overline t$ threshold \\
       &     &     &    &    & Scan $\stau2$ pair threshold \\ \hline
$\ee $ & 450 & L & 100 & 110 & Scan $\chz2~\chz3$ threshold \\ \hline
$\ee $ & 470 & L/R & 100 & 105 & Scan $\chpm1~\chmp2 $
                                threshold \\ \hline \hline
$e^-e^- $ & 265 & RR & 10 & 95 & Scan with $e^-e^-$ collisions for 
                   $\selr$ mass   \\
\end{tabular}
\end{ruledtabular}
\end{table*}

\begin{table*}
\caption{\label{tab:runsps1}
Run allocations for the SPS1 Minimal Sugra parameters. 
}
\begin{ruledtabular}
\begin{tabular}{cccccl}
Beams & Energy & Pol. & $\int {\cal L}dt$ 
  & [$\int {\cal L}dt]_{\rm equiv}$ & ~~~~~~~~Comments \\  \hline
$\ee $ & 500 & L/R & 335 & 335 & Sit at top energy for sparticle
       end point measurements \\ \hline
$\ee $ & 270 & L/R & 100 & 185 & Scan $\chz1~\chz2$ 
                                        threshold (R pol.) \\
       &     &     &     &     & Scan $\stau1~\stau1$ threshold 
                                        (L pol.) \\ \hline
$\ee $ & 285 & R & 50 & 85 & Scan $\smupr~\smumr$ threshold \\ \hline
$\ee $ & 350 & L/R & 40 & 60 & Scan $t\overline t$ threshold\\ 
       &     &     &     &     & Scan $\selr~\sell$ threshold 
                                      (L \& R pol.) \\
       &     &     &     &     & Scan $\chp1~\chm1$ threshold (L pol.) \\
                                        \hline
$\ee $ & 410 & L/R & 100 & 120 & Scan $\stau2~\stau2$ threshold \\ \hline
$\ee $ & 580 & L/R & 90 & 120 & Sit above $\chpm1~\chmp2$ threshold for 
	                           $\chpm2$ end point mass\\ \hline \hline
$e^-e^- $ & 285 & RR & 10 & 95 & Scan with $e^-e^-$ collisions for 
                   $\selr$ mass   \\
\end{tabular}
\end{ruledtabular}
\end{table*}

We should warn that a
rigorous optimization of the run scenarios has not been made, and
indeed the guidelines used for 
evaluating the two benchmark points are somewhat
different.  This study can be no more than an example that
a reasonable length run can provide good precision observables, since 
the multitude of possible SUSY (and Higgs) models still permitted 
is huge and the strategy and physics reach 
for each is likely to be quite different.

Tables~\ref{tab:runrr1} and \ref{tab:runsps1}
indicate both the luminosity allocated at a particular
energy, and the `equivalent luminosity' defined as that accumulation
that would have been made for the equivalent time spent at 500 GeV.
The 500 GeV equivalent luminosity is the appropriate
unit to account for the time spent.  It is the sum of the
equivalent luminosities that is set to 1 \iab ~in this study, 
and is the unit indicated in Table~\ref{tab:lumprofile}.

\section{Studies of supersymmetry}
\label{SUSY}

In this section we discuss the determination of the sparticle
mass precisions to be expected for the SPS1 benchmark run plan;
similar considerations were applied during the workshop to
the TESLA RR1 benchmark point and the results are simply summarized
here.   

The sparticle masses and dominant branching fractions
for the SPS1 point are given in Table~\ref{tab:sps1mass}.
The cross-sections at $\sqrt s = 500$ GeV for relevant two body processes in SPS1
are shown in Table~\ref{tab:sps1xs}.
Not shown in the Table are the corresponding squark, gluino and 
higgs masses and decay channels.  The lighter
squarks and gluino have masses $\sim 530$ and $595$ GeV respectively;
the two stop states have masses of 393 and 572 GeV.  The ($h^0, H^0,
A^0, H^\pm$) masses are respectively (113, 380, 379 and 388) GeV.

\begin{table*}
\caption{\label{tab:sps1mass}
Sparticle masses and dominant branching fractions for the SPS1
benchmark. 
}
\begin{ruledtabular}
\begin{tabular}{c|r|llllllll}
Particle & m(GeV)& ~~Final state /&(BR(\%)) & ~ & ~ & ~ & ~ & ~ & ~ \\ \hline
$\selr$ & 143 & $\chz1 e$ (100)~ \\
$\sell$ & 202 & $\chz1 e$ (45)~ & 
                  $\chpm1 \nu_e$ (34)~ &
                  $\chz2 e$ (20)~ \\
$\smur$ & 143 & $\chz1 \mu$ (100)~ \\
$\smul$ & 202 & $\chz1 \mu$ (45)~ & 
                  $\chpm1 \nu_\mu$ (34)~ &
                  $\chz2 \mu$ (20)~ \\
$\stau1$ & 135 & $\chz1 \tau$ (100)~ \\
$\stau2$ & 206 & $\chz1 \tau$ (49)~ & 
                  $\chm1 \nu_\tau$ (32)~ &
                  $\chz2 \tau$ (19)~ \\ \hline
$\snue$ & 186 & $\chz1 \nu_e$ (85)~ & 
                  $\chpm1 e^\mp$ (11)~ &
                  $\chz2 \nu_e$ (4)~ \\
$\snum$ & 186 & $\chz1 \nu_\mu$ (85)~ & 
                  $\chpm1 \mu^\mp$ (11)~ &
                  $\chz2 \nu_\mu$ (4)~ \\ 
$\snut$ & 185 & $\chz1 \nu_\tau$ (86)~ & 
                  $\chpm1 \tau^\mp$ (10)~ &
                  $\chz2 \nu_\tau$ (4)~ \\ \hline
$\chz1$ & 96 & stable~ \\
$\chz2$ & 175 & $\stau1 \tau$ (83)~ &
                  $\selr e$ (8)~ & 
                  $\smur \mu$ (8)~ \\
$\chz3$ & 343 & $\chpm1 W^\mp$ (59)~ & 
                  $\chz2 Z$ (21)~ & 
                  $\chz1 Z$ (12)~ & 
                  $\chz2 h$ (1)~  &
                  $\chz1 h$ (2)~ \\
$\chz4$ & 364 & $\chpm1 W^\mp$ (52)~ & 
                  $\snu \nu$ (17)~  &
                  $\stau2 \tau$ (3)~ &
                  $\chz1 Z$ (2)~ &
                  $\chz2 Z$ (2)~ & 
                  $\sell e$ (2)~ &
                  $\smul \mu$ (2)~ & 
                  $\widetilde\ell_R \ell$ (2)~ \\ \hline
$\chpm1$ & 175 & $\stau1 \tau$ (97)~ & 
                  $\chz1 q\overline q$ (2)~ & 
                  $\chz1 e\nu$ (0.4)~ &
                  $\chz1 \mu\nu$ (0.4)~ & 
                  $\chz1 \tau\nu$ (0.4)~   \\
$\chpm2$ & 364 & $\chz2 W$ (29)~ &
  	             $\chpm1 Z$ (24)~ &
                   $\widetilde\ell \nu_\ell$ (18)~ &
                   $\chpm1 h$ (15)~ &
                   $\snu_\ell \ell$ (8)~ &
                   $\chz1 W$ (6)~ \\
\end{tabular}
\end{ruledtabular}
\end{table*}

\begin{table}
\caption{\label{tab:sps1xs}
Selected cross sections in femtobarns 
for the SPS1 benchmark. Electron beam L and R
polarizations have magnitude 80\%.
Unless otherwise
noted, the energy is 500 GeV.
}
\begin{ruledtabular}
\begin{tabular}{c|cc||c|cc}
Reaction & $\sigma_L$ & $\sigma_R$ ~~ &
Reaction & $\sigma_L$ & $\sigma_R$ ~~ \\ \hline
$\chz1 \chz2$ & 105 & 25 ~~  &
$\selpl \selml$ & 105 & 17 ~~ \\
$\chz1 \chz3$ & 4 & 16 ~~ &
$\selpr \selmr$ & 81 & 546 ~~ \\
$\chz1 \chz4$ & 2 & 4 ~~ &
$\selpr \selml$ & 17 & 151 ~~ \\
$\chz2 \chz2$ & 139 & 16 ~~ &
$\selpl \selmr$ & 152 & 17 ~~ \\
$\chpm1 \chmp1$ & 310 & 36 ~~ &
$\smupr \smumr$ & 30 & 87 ~~ \\
$\chpm1 \chmp2$\footnotemark[1] & 7 & 2 ~~ &
$\smupl \smuml$ & 38 & 12 ~~ \\
$\chpm1 \chmp2$\footnotemark[2] & 37 & 10 ~~ &
$\stau1^+ \stau1^-$ & 35 & 88 ~~ \\
$\chpm1 \chmp2$\footnotemark[3] & 43 & 11 ~~ &
$\stau1^\pm \stau2^\mp$ & 2 & 1 ~~ \\
$\snue \snue^*$ & 929 & 115 ~~ &
$\stau2^+ \stau2^-$ & 31 & 11 ~~ \\
$\snum \snum^*$ & 18 & 14 ~~ & ~ & ~ & ~ \\
$\snut \snut^*$ & 18 & 14 ~~ & ~ & ~ & ~ \\ 

\end{tabular}
\end{ruledtabular}
\footnotetext[1] {For 540 GeV operation}
\footnotetext[2] {For 580 GeV operation}
\footnotetext[3] {For 620 GeV operation}
\end{table}

As indicated in Section~\ref{runplan}, the measurement of SUSY particle
masses relies on a first run at a high energy where
many sparticle pairs are produced.   In many of these cases,
the produced sparticles decay to two particles, one of which is a 
well-measured SM
particle, and the other is a sparticle (stable or unstable).
Since the original sparticles are mono-energetic (in the absence
of radiative losses from the incoming beams), their decay products
have flat energy distributions between lower and upper end points
fixed by the parent and decay sparticle masses.  Observation of these
end points then determines the masses of parent and decay sparticles.
The effects of initial state radiation and detector resolutions
will smear the energy distributions to some degree, but the end points 
can still be determined.  

\subsection{Energy end point mass measurements}

There have been several studies \cite{blairmartyn}\cite{colorado} 
of the precisions
obtainable in end point studies, 
incorporating smearing effects, for both the RR1
and SPS1 benchmark points, typically 
at 500 GeV.  Our estimates of end point mass
precisions are based upon simple \mbox{scaling} of statistical errors from
these studies to the number of events expected with our run plan.
This is perhaps an oversimplified model in the case where 
non-negligible backgrounds
for a particular process
from SM or other SUSY processes are present.  For the precisions to
be expected in the SPS1 benchmark, we have used, wherever
possible, the
results from Ref. \cite{colorado} which were done 
for the same SUSY scenario 
and thus have the appropriate SUSY backgrounds.

The $\selr$ and $\sell$ end point studies are particularly
rich, with distinct upper and lower edges coming from the distinct
$\selr$ and $\sell$ decays to $\chz1 e^\pm$.  The relative sizes and locations
 of these
end point edges in both $e^+$ and $e^-$ depend on the four
cross sections for $\selpr \selmr$, $\selpl \selml$,
$\selpl \selmr$ and $\selpr \selml$ for the 
different electron beam polarizations.
The energy spectra are further complicated by the presence
of \mbox{backgrounds} from the SM and SUSY processes, and by the fact that
the $\sell$  typically has other decays besides the
$\chz1 e$.
A new method to
facilitate these analyses by taking differences between
distributions with opposite beam polarizations, and between
emitted positron and electron distributions 
was developed in this workshop \cite{colselectron}.

The $\widetilde \tau$ and  $\chz2$ 
end point measurements 
studies are complicated by 
the fact that they decay dominantly into 
$\tau$ final states, and thus have missing 
neutrinos that wash out the energy end points.   
Nevertheless, the energy of the hadron from
$\tau$ 1-prong decays does carry some information on the
parent $\tau$ energy.  We have guessed, without direct
confirmation, that this may be sufficient to locate the
energy for the $\widetilde \tau$ and $\chz2$ 
threshold scans to within 1 -- 2 GeV.   

The $\chz3$ case is special for the SPS1 benchmark;
the $\chz3$ has an observable and distinctive decay
into $\chz1 Z$.  Using the well reconstructed $Z\rightarrow \ell^+
\ell^-$, the usual end point method works, albeit with low statistics.
The $\chz4$ is produced with insufficient rate at 500 GeV in this 
benchmark to be observable.

The charged states $\chpm1$ and $\chpm2$ pose special problems
also for end point measurements in the SPS1 benchmark.  
The dominant $\chpm1$ decay
is $\stau1 \nu_\tau$, which does not produce sharp end points.
However, $\snue \snue^*$ production with $\snue \rightarrow 
\chpm1 e^\mp$ is observable and permits a determination of
the $\chpm1$ mass.   The $\chpm2$ decay into $\chpm1 Z$ gives
a useful, but statistically limited, method for determining its
mass from the run at 580 GeV, above the $\chpm1 \chmp2$ threshold.
(Without some model assumptions, it is of course not possible
to know what energy is appropriate for the production 
of $\chpm1 \chmp2$, but the knowledge of the $\chpm1$ mass and
the measured $\snue$ pair cross-section, sensitive
to both $\widetilde \chi^\pm$ states, would give a good indication
of the $\chpm2$ mass.)

The end point analyses assume that it is possible to find a final 
state that can be clearly identified as arising from a particular two
body reaction.  This assumption needs to be examined 
carefully, as in practice many two-body processes can feed the same final
state.  The details vary strongly with benchmark point.  

\begin{table*}
\caption{\label{tab:finalstates}
The dominant contributors to some specific final states, with
specified initial electron beam polarization.  $N$ is the 
number events expected (before acceptance and efficiency cuts)
in the 335 \ifb ~allocated to the 500 GeV run.  The percentage
of these events from the dominant reaction is $\cal F$. The L and
R $e^-$ beam polarizations were taken with magnitude 80\%.
}
\begin{ruledtabular}
\begin{tabular}{c|cccccc}
Final state & Pol & $N$ & dominant & $\cal F$ & SM particles & masses \\
 ~      & ($e^-$) & ~ & reaction(s) &  ~     & used         & measured \\
          \hline
$e^+e^-$\me & R/L & 210K/65K & 
     $\sell \sell,~\selr \selr,~\sell \selr$ & 92 
     & $e^\pm$ & $\sell, \selr, \chz1$ \\
$\mu^+\mu^-$\me & 
  R & 31K & $\smur \smur$ & 95 & $\mu^\pm$ & $\smur, \chz1$ \\
$\tau^+\tau^-$\me & 
  L & 152K & $\chpm1 \chmp1$ & 64 & $\tau^\pm$ & $\chpm1, \stau1$ \\
$e^\pm \tau^\mp$\me & 
  L & 88K & $\snue \snue^*$ & 65 & $e^\pm$ & $\chpm1, \snue$ \\
$\mu^+\mu^-\tau^+\tau^-$\me & 
  L & 2K & $\smul \smul$ & 97 & $\mu^\pm$ & $\smul, \chz1, \chz2$ \\
$e^+e^-\tau^+\tau^-$\me & 
  R & 10K & $\sell \selr$ & 91 & $e^\pm$ & $\sell, \chz1, \selr$ \\
$\tau^+\tau^-\tau^\pm\mu^\mp$\me & 
  R & 8K & $\snum \snum^*$ ($\smul \smul$) & 
43 (57) & $\mu^\pm$ & $\snum, \chpm1$ \\
\end{tabular}
\end{ruledtabular}
\end{table*}

We have looked at the competing reactions feeding
particular final states of 2 or 4 leptons ($e, \mu, \tau$) plus 
missing energy (\me ) for the SPS1 benchmark 
\cite{caveat}.
We have required here that the final states contain no
strongly interacting particles.
For example, after taking all the cross sections and 
branching ratios into account,  the 
contributions to the $e^\pm \tau^\mp$\me ~final state
with right polarized electron beam are 
spread over the initial channels:
$\sell \sell$ (5); $\selr \sell$ (56); $\chp1 \chm1$ (0.3);
$\snue \snue^*$ (21); and $\chz1 \chz3$ (0.8)
(where the numbers in parentheses are $\sigma \times$BR in fb).  
For such cases with multiple competing reactions, 
attributing structures in the energy distributions
to particular SUSY particles will be difficult.  In general,
the end point analyses are likely to require iterative
approaches to separate effects of the different sparticles.  

Nevertheless, for
benchmark SPS1, we find that most sleptons and gauginos can
be reasonably well isolated in specific channels.  The
$\selr$ and $\sell$ are mixed in the $e^+e^-$\me ~final states, 
but Ref. \cite{colselectron}
shows that they can be disentangled.  
Table~\ref{tab:finalstates} shows some of the final states
that are dominated by specific sparticle production processes.
We see that apart from the $\snut,\chz2$ and perhaps
$\snum$, there is at least one process that allows relatively
clean access to each of the sparticle masses through end point
measurements.  (Recall that the $\chz3$ mass can be accessed
in the SPS1 benchmark through its decay into $\chz1 Z$).

We caution however that, although for the SPS1 case examined
here one can find channels that are specifically sensitive
to particular sparticle masses, it is by no means clear that
that one will easily deduce the sparticles responsible
for the observed end points when one does not {\it a priori} know the
SUSY model.

Table~\ref{tab:measmass} shows our estimates of the precisions
obtainable by end point measurements for the SPS1 benchmark, based
on the run plan shown in Table \ref{tab:runsps1}.  The caveats of
the preceding paragraphs suggest these should be taken only as
educated guesses; a complete Monte Carlo calculation including
all SUSY processes and SM backgrounds should be made.

\subsection{Threshold scan mass measurements}

Once one has an estimate of sparticle masses from the end point 
measurements, refined \mbox{determinations} can often be obtained
by performing a scan across the threshold of a reaction involving
that sparticle.  For these studies, it is not necessary to
restrict attention to easily reconstructed final states; it is
sufficient that the final states are observable in the detector
and that other thresholds in the same polarization and final
\mbox{state} do not occur in the same region.
Studies of such threshold scans have been made in Ref. 
\cite{blairmartyn} for benchmark RR1.   In that study, 100 \ifb 
~were devoted to each scan, with runs distributed over 10 equidistant
energy points.   This strategy is almost surely not ideal; an
optimized scan algorithm should depend upon the amount of background
in the channels observed, the total cross-section times branching
ratio, the uncertainty in $\sigma\times$BR, and on the steepness of the
threshold curve as a function of energy.  
Ref \cite{kenichi}
has studied an optimization for $\snum$ and $\snut$ thresholds where the
cross sections are small and find that two points on the rise
of the cross-section and one well above threshold are more suitable.
A study performed for this workshop \cite{blairthresh}
has investigated how to obtain both sparticle
masses and total widths, and finds that a two point scan may be
optimum.  This analysis also concludes that the widths 
for many of the states may be accessible at the 35 -- 50\% level.

As part of this study, we have made an analytic estimate
of the accuracy available
in a threshold scan \cite{cahn} for the case that 
equal luminosities are collected at $N$ scan points,
spaced at equal energy, $\delta E$.  The threshold is assumed to be
within $\delta E$ below the first of the scan
points.  No background is included in these studies. 
The presence of beamstrahlung should not affect the threshold
turn-on markedly, since the collisions at the
dominant peak at the full beam energy give an unsmeared
threshold behavior.
Minimizing the likelihood function formed
from the Poisson probabilities to give the observed
numbers of events at each energy point, we can determine the
most likely value of the threshold energy and hence sparticle
mass.   The analytic results can be approximated as: 
$$ \delta m \approx \Delta E (1 + 0.36/\sqrt N)/
   \sqrt{18N{\cal L}\sigma_u} $$
for a $\beta^3$ $p$-wave threshold, and
$$ \delta m \approx \Delta E N^{-1/4}(1 + 0.38/\sqrt N)/
   \sqrt{2.6N{\cal L}\sigma_u} $$
for a $\beta^1$ $s$-wave threshold.  
Here, $\Delta E$ is the full energy interval over which the scan is made,
$\cal L$ is the total luminosity devoted to each point of
the scan,
$N$ is the number of energy settings,
and $\sigma_u$ is the cross-section at the upper energy of
the scan.  Note that the $p$-wave threshold benefits
little from increasing the number of energy settings
above 3 to 4, while an \mbox{$s$-wave} threshold precision
continues to improve weakly as $N^{-1/4}$ with the number of points
in the scan.
These analytic approximations are in good agreement with the 
Monte Carlo precisions for the $p$-wave $\ee \rightarrow
\smupr \smumr$ and $s$-wave $\ee \rightarrow \chp1 \chm1$ 
threshold scans of Ref.~\cite{blairmartyn}.  

The run plans for the RR1 and SPS1 benchmark points call for
scans as indicated in Tables \ref{tab:runrr1} and \ref{tab:runsps1}.
In both, a special scan at the $\selr$ threshold is called for
using right polarized $e^-$ beams; this strategy \cite{fengpeskin}
is dictated by the fact that the $\selr \selr$ threshold energy
cross section rises as $\beta$ in $e^-e^-$, whereas in $\ee$ it
rises as $\beta^3$.  The sharper rise, even after inclusion
of the effects of beamsstrahlung gives a better determination
of the $\selr$ mass.  

In the RR1 scenario, the $\snue , \snum, \snut$ states are 
observable through their decays into $\chpm1 \ell^\mp$
with subsequent $\chpm1 \rightarrow  \chz1 q \overline q$ and 
$\chpm1 \rightarrow \chz1 \ell \nu$, although the \mbox{event} 
rates are small.
We thus include a scan at the $\snu$ pair threshold around 320 GeV
to get some mass information, estimated on the basis
of the analysis in Ref \cite{kenichi}.  (The more precise
$\snu$ mass determination in Ref \cite{blairmartyn} seems
to be too optimistic for this channel.)
In the SPS1 benchmark, the
$\chpm1$ decays dominantly into $\stau1 \nu_\tau$ and the
signature is hard to dig out from background.  In the SPS1
case, we thus do not call for a $\snu$ pair threshold scan.

We have estimated sparticle mass precisions 
from threshold scans by simple statistical scaling of the results
of Ref \cite{blairmartyn}, based on the ratio of 
$\sigma\times{\rm BR}\times{\cal L}$
for the appropriate reaction in our run scenario to
that used in Ref \cite{blairmartyn}.  We use the 
reaction cross-sections at 500 GeV for this scaling.
This simple estimating procedure is doubtless too naive,
since it ignores details of the backgrounds at different
benchmark points, and has not incorporated the effects of
uncertainties in the knowledge of $\sigma\times{\rm BR}$.

The resulting estimates of the mass precisions from the scans
in benchmark SPS1 and run plan of Table \ref{tab:runsps1}
are given in Table \ref{tab:measmass},
together with the combination in quadrature for
the end point and scan mass errors, where both are available.

Similar mass error estimates 
for the benchmark \mbox{RR1}, worked out in less detail
at the Workshop for the run plan in Table \ref{tab:runrr1},
are also given in Table \ref{tab:measmass}.   In general, we expect
that the precisions for the RR1 case will be better than
for SPS1, owing to the smaller sparticle masses (and higher
cross-sections), and to the smaller decay branching ratios
into $\tau$'s.   Specific differences between
any two benchmarks always exist.  
The decay $\chz3 \rightarrow \chz1 Z$, open in the SPS1
case but not the RR1 case, is illustrative of this. 

\begin{table}
\caption{\label{tab:measmass}
Mass precision estimates in GeV for benchmark point SPS1 
for end point (EP), threshold scan (TH) and combined 
measurements, and the combined estimates for the RR1 point.}
\begin{ruledtabular}
\begin{tabular}{c|ccc|c}
 ~ & ~ & SPS1 & ~ & RR1 \\
particle & $\delta M_{\rm EP}$ &
           $\delta M_{\rm TH}$ & $\delta M_{\rm SPS1}$ & 
	     $\delta M_{\rm RR1}$ \\ \hline
$\selr$ & 0.19 & 0.02 & 0.02 & 0.02 \\
$\sell$ & 0.27 & 0.30  & 0.20 & 0.20 \\
$\smur$ & 0.08 & 0.13  & 0.07 & 0.13 \\
$\smul$ & 0.70 & 0.76  & 0.51 & 0.30 \\
$\stau1$ & $\sim 1 - 2$ & 0.64  & 0.64 & 0.85 \\
$\stau2$ & -- & 0.86  & 0.86 & 1.34 \\
$\snue$ & 0.23 & --  & 0.23 & 0.4 \\
$\snum$ & 7.0 & --  & 7.0 & 0.5 \\
$\snut$ & -- & --  & -- & 10.0 \\
$\chz1$ & 0.07 & --  & 0.07 & 0.07 \\
$\chz2$ & $\sim 1 - 2$ & 0.12  & 0.12 & 0.30 \\
$\chz3$ & 8.5 & --  & 8.5  & 0.30 \\
$\chz4$ & -- & --  & -- & observed \\
$\chpm1$ & 0.19 & 0.18  & 0.13 & 0.09 \\
$\chpm2$ & 4.1 & --  & 4.1 & 0.25 \\
\end{tabular}
\end{ruledtabular}
\end{table}

\subsection{SUSY model parameter determination}
 
Once we have measured sparticle masses, we will want
to estimate the underlying supersymmetry
parameters, and to probe the character of the SUSY-breaking.
We have noted above that in general it will be a challenge 
to determine the nature of the SUSY-breaking model, but
the totality of information from \lhc ~and \lc ~should
give us good indicators.  The recent work \cite{blairporodzerwas}
analyzing  the renormalization group evolution of masses
suggests that at least it is possible to cleanly distinguish
the class of SUSY model (e.g. mSUGRA {\it vs.} gauge mediated
SUSY).

It is a separate matter
to ask, given the hypothesis of the SUSY model, how
well its parameters may be determined.
For the two SUSY points considered here, we have
made an estimate of the precision on the underlying
SUSY parameters assuming that mSUGRA is at work.
For the mSUGRA scenarios considered here, we expect that the 
errors on $m_0$ and $m_{1/2}$ are mainly determined by the
errors on the ($\selr, \smur$) and ($\chpm1, \chpm2$) 
masses respectively.
The errors on $A_0$ and $\tan\beta$ should be primarily
controlled by the errors on ($\stau1, \stau2$) and 
($\chpm1 ,\chz1$) masses respectively.  

We use the full set of mass error estimates of Table~\ref{tab:measmass}
for the SPS1 benchmark point and propagate them to give the mSUGRA parameter
errors.  These agree well with those given in \cite{blairmartyn} by 
the above simplified relations, after scaling for the number of
observed events.
The resulting mSUGRA parameter estimates are given
in Table~\ref{tab:msugraerr}.  These estimates are 
conservative since they do not include potential
information from stop masses, nor from the heavier Higgs
sector and these may be expected to help materially.
Similarly, information on the polarized cross-sections
should help to further constrain $A_0$.

\begin{table}
\caption{\label{tab:msugraerr}
Errors on mSUGRA mass parameters for the SPS1 and RR1 
hypotheses.}
\begin{ruledtabular}
\begin{tabular}{c|c|c}
parameter & SPS1 & RR1 \\ \hline
$m_0$ & $100 \pm 0.08$ GeV ~~~ & $100 \pm 0.04$ GeV ~~~  \\
$m_{1/2}$ & $250 \pm 0.20$ GeV ~~~ & $200 \pm 0.22$ GeV ~~~ \\
$A_0$ & $0 \pm 13$ GeV ~~~ & $0 \pm 18$ GeV ~~~ \\
$\tan\beta$ & $10 \pm 0.47$ ~~~ & $3 \pm 0.05$ ~~~ \\
\end{tabular}
\end{ruledtabular}
\end{table}

\section{Higgs boson}
\label{higgs}

The Higgs boson properties should be determined with as
high accuracy as possible to seek departures from the SM
and constrain the parameters of potential new physics models.
Previous studies \cite{teslarr1}\cite{orange} have 
estimated the errors on the Higgs mass, cross-sections,
total and partial widths, and branching ratios for $m_H=120$ GeV.
These studies have used the Higgs bosons produced in the
reaction $\ee \rightarrow ZH$ only.  They use multivariate
analyses based on information from jet topology and
separated vertex information to extract the fermionic
branching ratios statistically.   Combination of the
cross-sections for $ZH$ and $\nu\nu H$ ($WW$ fusion) and total
width measurements allows the determination of the bosonic
couplings.
   
Using the $ZH$ cross-section as a function of $\sqrt s$ from Pythia,
and the scenarios proposed in Section~\ref{runplan}, we
find that our run plans produce as many $ZH$ as would
be obtained in dedicated running at 350 GeV with 550 (650) \ifb
or with 1280 (1350) \ifb at 500 GeV for the run plans of Table~\ref{tab:runsps1}
(Table~\ref{tab:runrr1}).  
We do not expect that operation of the collider at several
different energies, as envisioned in our run plans, will
materially degrade the Higgs studies, as it is mostly just
the number of $ZH$ events that matters.
In Table \ref{tab:higgs} we show the estimated Higgs parameter
errors obtained by \mbox{statistical} scaling from the number of $ZH$ events
in the run scenario of Table~\ref{tab:runsps1}.  The errors for the run
plan of Table~\ref{tab:runrr1} are about 10\% better.

We note that the top quark cross section
near threshold
depends upon the $t\overline t H$ Yukawa coupling.
Ref \cite{orange} indicates that a 14\% variation of top 
Yukawa coupling results in a 2\% change in $\sigma_{t\overline t}$.
However, if \lc ~operation above the $t\overline t H$
threshold is possible, direct measurement of the cross-section will
give an improved precision.  References \cite{teslarr1}\cite{orange} 
indicate that 1000 \ifb ~at $\sqrt s=800$ GeV will result in
$\delta\lambda_{t\overline t H} = 5.5$\% for $m_H=120$ GeV, degrading
to about 25\% at 500 GeV.

\begin{table}
\caption{\label{tab:higgs}
Relative errors (in \%) on Higgs mass, cross-section,
total width, branching ratios and Yukawa couplings ($\lambda$)
for the run plan of Table \ref{tab:runsps1}.
}
\begin{ruledtabular}
\begin{tabular}{cc|cc}
Parameter & error & Parameter & error \\ \hline
Mass    & 0.03       & 
       $\Gamma_{\rm tot}$ & 7 \\
$\sigma$($ZH$) &  3  & 
       $\lambda_{ZZH}$   & 1     \\
$\sigma$($WW$) & 3   & 
       $\lambda_{WWH}$   & 1     \\
BR($b\overline b$) & 2 & 
       $\lambda_{bbH}$   & 2      \\
BR($c\overline c$) & 8 & 
       $\lambda_{ccH}$   & 4      \\
BR($\tau^+\tau^-$) & 5 & 
       $\lambda_{\tau\tau H}$ & 2      \\
BR($gg$)          & 5 & 
       $\lambda_{ttH}$   & 30    \\
\end{tabular}
\end{ruledtabular}
\end{table}

\section{Top Quark}
\label{top}

The top quark parameters are determined from the scan near
the $t\overline t$ threshold at 350 GeV.  The statistical
errors \cite{teslarr1} \cite{orange} 
are small compared to the uncertainties 
in the theoretical errors arising from the QCD theory.

The top quark mass parameter may be defined in several ways; the
pole mass used in the Tevatron experimental studies is uncertain
at the level of 0.5 GeV due to non-perturbative 
renormalon effects.  If one uses alternate mass definitions,
such as one half of the toponium quasi-bound state mass, the
non-perturbative effects are reduced.

The top quark width can be determined from the $t\overline t$ 
threshold scan since the cross-section at the 1S bound state
energy is proportional to $1/\Gamma_t$.   Added information
on the width can be obtained from the forward-backward asymmetry
which is non-zero due to interference of diagrams involving
the $t\overline t \gamma$, $t\overline t Z$ and $t\overline t H$ couplings.

Ref. \cite{teslarr1} \cite{orange} suggest
that the top quark mass should be measured with an error
of 150 MeV.  The width, expected to be about 1.4 GeV in the SM, 
should be determined to within 5\%.

\section{Summary}
\label{summary}

We have examined how a Linear Collider program of 1000 \ifb could
be constructed in the case that a very rich program of new
physics is accessible at $\sqrt s \le 500$ GeV.  We have examined
possible run plans that would allow the measurement of the 
parameters of a 120 GeV Higgs boson, the top quark,
and could give information
on the sparticle masses in SUSY scenarios in which many 
states are accessible.

We find that the construction of the run plan (the specific
energies for collider operation, the mix of initial 
state electron polarization states, and the use of special
$e^-e^-$ runs) will depend quite sensitively on the 
specifics of the supersymmetry model, as the decay channels
open to particular sparticles vary drastically and 
discontinuously as the 
underlying SUSY model parameters are varied.  We have explored
this dependence somewhat by considering two rather closely
related SUSY model points.  We have called for operation
at a high energy to study kinematic end points, followed by runs
in the vicinity of several \mbox{two} body production thresholds
once their location is determined
by the end point studies.  For our benchmarks, the end point
runs are capable of disentangling most sparticle states
through the use of specific final states and
beam polarizations.  The estimated sparticle mass precisions,
combined from end point and scan data, 
are given in Table~\ref{tab:measmass} and the corresponding
estimates for the mSUGRA parameters are in Table~\ref{tab:msugraerr}.

The precision for the Higgs boson mass, width, cross-sections,
branching ratios and couplings are given in Table~\ref{tab:higgs}.
The errors on the top quark mass and width are expected to be
dominated by the systematic limits imposed by QCD non-perturbative
effects.

The run plan devotes at least two thirds of the accumulated luminosity
near the maximum \lc ~energy, so that the program would be 
sensitive to unexpected new phenomena at high mass scales.

We conclude that with a 1 \iab ~program, expected to take the first
6 -- 7 years of \lc ~operation, one can do an excellent job of providing
high precision measurements with which to probe the nature of
the new physics, and which will give complementary and improved 
information over that obtained at the \lhc .

%
%



\begin{thebibliography}{99}
\bibitem{caveat}  This study was done by hand; the detailed results should
be checked by more careful Monte Carlo analysis.
\end{thebibliography}

\end{document}
%